\documentclass[a4paper]{jpconf}
\usepackage{graphicx,cite}
\begin{document}
\title{Hybrid classical integrable structure of squashed sigma models - {\it a short summary} - }

\author{Io Kawaguchi and Kentaroh Yoshida}

\address{Department of Physics, Kyoto University Kyoto 606-8502, Japan}

\ead{io@gauge.scphys.kyoto-u.ac.jp, kyoshida@gauge.scphys.kyoto-u.ac.jp}

\begin{abstract}
We give a short summary of our recent works on the classical integrable structure of 
two-dimensional non-linear sigma models defined on squashed three-dimensional spheres. 
There are two descriptions to describe the classical dynamics, 1) the rational description 
and 2) the trigonometric description. 
It is possible to construct two different types of Lax pairs depending on the descriptions, and 
the classical integrability is shown by computing classical $r/s$-matrices satisfying the extended Yang-Baxter equation 
in both descriptions. In the former the system is described as an integrable system of rational type. On the other hand, 
in the latter it is described as trigonometric type. 
There exists a non-local map between the two descriptions 
and those are equivalent. This is a non-local generalization of the left-right duality in principal chiral models.  
\end{abstract}

\section{Introduction}

The notion of integrability is of significance in theoretical and mathematical physics. 
In integrable quantum field theories it may be possible to prove strong-weak dualities, 
though it is quite difficult to prove it in general. 
It is well recognized that integrability plays an important role toward the proof of AdS/CFT \cite{Maldacena} 
after the recent progress (For a comprehensive overview see \cite{review}). 
The symmetric coset structure of AdS spaces and spheres is a key ingredient \cite{BPR}. 
The classification of symmetric cosets, which are potentially applicable in the context of AdS/CFT, 
has also been performed in \cite{Zarembo}.

It is quite natural to consider integrable deformations of AdS spaces and spheres to figure out a larger integrable structure 
behind the gauge/gravity (string) correspondence. 
One of the most investigated examples 
is the Lunin-Maldacena background \cite{LM} and its gauge-theory counter part \cite{LS,Berenstein}. 
Then the Lax pair for strings on this background is discussed \cite{Frolov}. 
Motivated by this work, we will consider another kind of integrable deformation of spheres, {\it squashed spheres}. 

We focus upon the classical integrable structure of two-dimensional non-linear sigma models 
defined on three-dimensional squashed spheres. Since the squashed spheres are described as non-symmetric cosets, 
the integrability is not so obvious in comparison to the symmetric case such as principal chiral models and $O(N)$ non-linear sigma models \cite{Luscher2}. 

In this short summary we will summarize the results on the classical integrable structure of 
two-dimensional non-linear sigma models defined on squashed three-dimensional spheres, especially focusing on \cite{KYhybrid}. 
There are two descriptions to describe the classical dynamics, 1) the rational description 
and 2) the trigonometric description. 
It is possible to construct two different types of Lax pairs depending on the descriptions, and 
the classical integrability is shown by computing classical $r/s$-matrices satisfying the extended Yang-Baxter equation 
in both descriptions. In the former the system is described as an integrable system of rational type. On the other hand, 
in the latter it is described as trigonometric type. 
There exists a non-local map between the two descriptions 
and those are equivalent. This is a non-local generalization of the left-right duality in principal chiral models.  
A similar analysis here is applicable to non-linear sigma models defined on three-dimensional Schr{\"o}dinger spacetimes 
\cite{KYSch}.

The content is the following. In section 2 we introduce three-dimensional squashed spheres and 
the classical action of two-dimensional sigma models defined on three-dimensional squashed spheres.  
In section 3 we explain the rational description based on $SU(2)_{\rm L}$ to describe the classical dynamics. 
A Lax pair is constructed with an improved $SU(2)_{\rm L}$ current. The $r/s$-matrix algebra is computed and 
the classical integrability is shown as a rational model.  
In section 4 the classical integrability is discussed in the trigonometric description based on $U(1)_{\rm R}$\,. 
The $r/s$-matrix algebra implies that the system is described as a trigonometric model.  
There is a non-local map between the two descriptions and so they are equivalent.  
Section 5 is devoted to conclusion and discussion.

\section{Setup}

Three-dimensional squashed spheres are described as one-parameter deformations of round sphere $S^3$ and the metric is given by 
\begin{eqnarray}
ds^{2}=\frac{L^{2}}{4}\left[d\theta^{2}+\sin^{2}\theta d\phi^{2}+(1+C)\left(d\psi+\cos\theta d\phi\right)^{2}\right]\,.
\label{squashed angle}
\end{eqnarray}
The constant $C$ is a deformation parameter. For $C=0$, the metric (\ref{squashed angle}) is reduced to that of round S$^{3}$ 
with radius $L$. For $C=-1$, the squashed spheres shrink to round S$^{2}$ with radius $L/2$. 
The metric (\ref{squashed angle}) is invariant under the $SU(2)_{\rm L}\times U(1)_{\rm R}$ transformations 
\begin{eqnarray}
\delta^{L,1}\bigl(\phi,\theta,\psi\bigr)&=&\left(-\frac{\cos\phi}{\tan\theta},-\sin\phi,~\frac{\cos\phi}{\sin\theta}\right)\,, \nonumber \\
\delta^{L,2}\bigl(\phi,\theta,\psi\bigr)&=&\left(-\frac{\sin\phi}{\tan\theta},~\cos\phi,~\frac{\sin\phi}{\sin\theta}\right)\,, \nonumber \\
\delta^{L,3}\bigl(\phi,\theta,\psi\bigr)&=&\left(1,~0,~0\right)\,, \nonumber \\
\delta^{R,3}\bigl(\phi,\theta,\psi\bigr)&=&\left(0,~0,-1\right)\,. \nonumber 
\end{eqnarray}

It is convenient to rewrite the metric (\ref{squashed angle}) in terms of an $SU(2)$ group element 
represented by 
\begin{eqnarray}
g={\rm e}^{T^{3}\phi}{\rm e}^{T^{2}\theta}{\rm e}^{T^{3}\psi}\,,
\end{eqnarray}
where $T^{a}~(a=1,2,3)$ are the generators of $SU(2)$ and satisfy the following relations, 
\begin{eqnarray}
\left[T^{a},T^{b}\right] = \varepsilon^{ab}_{~~c}T^{c}\,, \quad 
{\rm Tr}\left(T^{a}T^{b}\right) = -\frac{1}{2}\delta^{ab}\,, \nonumber 
\end{eqnarray}
and $\varepsilon^{ab}_{~~c}$ is the totally antisymmetric tensor normalized $\varepsilon_{123}=+1$. 
The group indices are raised and lowered by using the Killing two-form $\delta_{ab}$. 

By using the left-invariant one-form 
\[
J \equiv g^{-1}dg\,, \qquad g \in SU(2)\,,
\]
the metric (\ref{squashed angle}) can be rewritten as 
\begin{eqnarray}
ds^{2} 
&=&-\frac{L^{2}}{2}\left[
{\rm Tr}\left(J^{2}\right)
-2C\left({\rm Tr}\left[T^{3}J\right]\right)^{2} \right]\,.
\label{squashed}
\end{eqnarray}
The $SU(2)_{\rm L}$ transformation is just the left action and
$U(1)_{\rm R}$ transformation is the right action generated by $T^{3}$\,, 
\begin{eqnarray}
g \rightarrow g^L \cdot g \cdot {\rm e}^{-T^{3}\alpha}\,.
\end{eqnarray}
The infinitesimal forms are 
\begin{eqnarray}
\delta^{L,a}g = \epsilon\,T^{a}g\,, \qquad \delta^{R,3}g = -\epsilon\,gT^{3}\,. 
\label{trans-1}
\end{eqnarray}

Let us consider two-dimensional non-linear sigma models defined on squashed spheres with the metric (\ref{squashed}). 
These sigma models are called ``squashed sigma models'' as an abbreviation.  
The classical action is given by 
\begin{eqnarray}
S =\!\! \int\!\!\!\!\int\!\!dtdx\Bigl[
{\rm Tr}\left(J_{\mu} J^{\mu}\right)
-2C\,{\rm Tr}\!\left(T^{3}J_{\mu}\right)\!{\rm Tr}\!\left(T^{3}J^{\mu}\right)
\Bigr]. 
\label{sigma}
\end{eqnarray}
The coordinates and metric of base space are $x^{\mu}=(t,x)$ and $\eta_{\mu\nu}={\rm diag}(-1,+1)$\,. 
Suppose that the value of $C$ is restricted to $C > -1$ 
so that the sign of kinetic term is not flipped. 
The action (\ref{sigma}) is invariant under (\ref{trans-1}). 

Note that the Virasoro conditions and periodic boundary conditions are not imposed here, although we are potentially interested 
in applications to string theory. 
Instead, we impose the boundary condition that the group variable $g(x)$ approaches very rapidly a constant element 
as it goes to spatial infinities like 
\begin{eqnarray}
g(x) \to g_{(\pm)}~:~\mbox{const.} \qquad (x \to \pm \infty)\,. \label{bc}
\end{eqnarray}
Thus the left invariant one-form $J_{\mu}(x)$ vanishes as it approaches spatial infinities. 

The equations of motion are 
\begin{eqnarray}
&& \partial^{\mu}J_{\mu} - 2C {\rm Tr}(T^3\partial^{\mu}J_{\mu})T^3 
- 2C\,{\rm Tr}(T^3J_{\mu})[J^{\mu},T^3] =0\,.
\label{eom}
\end{eqnarray}
Multiplying $T^a$ and taking the trace, we obtain the $T^a$ component of (\ref{eom}). 
The $T^3$ component is nothing but the conservation law of the $U(1)_{\rm R}$ current, 
\begin{eqnarray}
(1+C)\partial^{\mu}J^{3}_{\mu} = 0\,.
\label{eom3}
\end{eqnarray}
The $T^\pm$ components are 
\begin{eqnarray}
\partial^{\mu}J^{\pm}_{\mu} \mp iCJ^{3,\mu}J^{\pm}_{\mu} = 0\,, 
\label{eompm}
\end{eqnarray}
where $T^{\pm}$ are defined as 
\begin{eqnarray}
T^{\pm}\equiv\frac{1}{\sqrt{2}}\left(T^{1}\pm iT^{2}\right)\,.
\end{eqnarray}
The equations of motion (\ref{eom}) are equivalent to a set of (\ref{eom3}) and (\ref{eompm}). 
Note that the equations of motion (\ref{eom}) are equivalent to the conservation law of the $SU(2)_{\rm L}$ current,  
\begin{eqnarray}
\partial^{\mu}\left[gJ_{\mu}g^{-1}-2C{\rm Tr}\left(T^{3}J_{\mu}\right)gT^{3}g^{-1}\right]=0\,.
\end{eqnarray}

Hereafter we will see that there are the two descriptions to describe the classical dynamics,  
1) the rational description based on $SU(2)_{\rm L}$ and 2) the trigonometric one based on $U(1)_{\rm R}$\,. 
In each of the descriptions, it is possible to construct a different Lax pair which leads to the identical equations of motion 
(\ref{eom}). 

\section{Rational description}

Let us consider a description based on $SU(2)_{\rm L}$\,. By taking account of the ambiguity of the Noether current, 
the $SU(2)_{\rm L}$ conserved current is generally written as  
\begin{eqnarray}
j_{\mu}^{L} \equiv g J_{\mu} g^{-1}-2C{\rm Tr}\left(T^{3}J_{\mu}\right)gT^{3}g^{-1}+\epsilon_{\mu\nu}\partial^{\nu}f\,,
\end{eqnarray}
where $f$ is an arbitrary function and the antisymmetric tensor $\epsilon_{\mu\nu}$ on the world-sheet 
is normalized $\epsilon_{tx}=+1$\,. The first two terms are obtained by the standard procedure. 
The last term is a topological term and it can be taken freely under the boundary condition (\ref{bc}). 

When $f$ is taken as 
\begin{eqnarray}
f=-\sqrt{C}gT^{3}g^{-1}\,, \label{f}
\end{eqnarray}
the $SU(2)_{\rm L}$ current $j^{L}_{\mu}$ is improved so as to satisfy the flatness condition \cite{KY}: 
\begin{eqnarray}
\epsilon^{\mu\nu}\left(\partial_{\mu}j^{L}_{\nu}-j^{L}_{\mu}j^{L}_{\nu}\right)=0\,. \label{flat}
\end{eqnarray}
The current improvement leads to a deformed current algebra given by 
\begin{eqnarray}
\left\{j^{L,a}_{t}(x),j^{L,b}_{t}(y)\right\}_{\rm P} 
&=& \varepsilon^{ab}_{~~c}j^{L,c}_{t}(x)\delta(x-y)\,, \nonumber \\
\left\{j^{L,a}_{t}(x),j^{L,b}_{x}(y)\right\}_{\rm P} 
&=& \varepsilon^{ab}_{~~c}j^{L,c}_{x}(x)\delta(x-y)+(1+C)\delta^{ab}\partial_{x}\delta(x-y)\,, \nonumber \\
\left\{j^{L,a}_{x}(x),j^{L,b}_{x}(y)\right\}_{\rm P} 
&=& -C\varepsilon^{ab}_{~~c}j^{L,c}_{t}(x)\delta(x-y)\,, \nonumber 
\end{eqnarray} 
where $j_{\mu}^{L,a} \equiv -2{\rm Tr}(T^{a} j_{\mu}^{L})$\,. The last bracket does not vanish due to the improvement. 

Then, by using the flat $SU(2)_{\rm L}$ current with (\ref{f})\,, a Lax pair is constructed as  
\begin{eqnarray}
L^{L}_{t}(x;\lambda) = \frac{1}{1-\lambda^{2}}\left[j^{L}_{t}(x)-\lambda j^{L}_{x}(x)\right]\,, \quad 
L^{L}_{x}(x;\lambda) = \frac{1}{1-\lambda^{2}}\left[j^{L}_{x}(x)-\lambda j^{L}_{t}(x)\right]\,, 
\label{left lax}
\end{eqnarray}
where the $\lambda$ is a spectral parameter. 
The commutation relation
\begin{eqnarray}
\left[\partial_{t}-L_{t}^L(\lambda),
\partial_{x}- L_{x}^L(\lambda)\right] = 0 
\end{eqnarray}
leads to the equations of motion (\ref{eom}) and the flatness condition (\ref{flat}). 

With the Lax pair (\ref{left lax}), the monodromy matrix $U^L(\lambda)$ is defined as 
\begin{eqnarray}
U^L(\lambda) \equiv {\rm P}\exp{\left[\int^{\infty}_{-\infty}\!\!dx\,L_{x}^L(x;\lambda)\right]}\,. \nonumber
\end{eqnarray}
The symbol P means the path ordering. 
Since the Lax pair can be regarded as a flat connection, the monodromy matrix $U^L(\lambda)$ is conserved, 
\begin{eqnarray}
\frac{d}{dt}U^L(\lambda)=0\,. \nonumber
\end{eqnarray}
Thus expanding it around a fixed value of $\lambda$ can generate an infinite number of conserved charges.  
The expression of the charges depends on the expansion point around which the monodromy matrix has been expanded. 
The expansion around $\lambda=\infty$ leads to an infinite number of the non-local charges constructed in \cite{KY}.  
When it is expanded around $\lambda=\pm 1$\,, an infinite number of commuting local charges (in involution) 
which ensure the classical integrability in the sense of Liouville. 

The Poisson bracket of $L^{L,a}_{x}(x;\lambda)$ is evaluated as 
\begin{eqnarray}
\left\{L^{L,a}_{x}(x;\lambda), L^{L,b}_{x}(y;\mu)\right\}_{\rm P}&=&\frac{1}{\lambda-\mu}\varepsilon^{ab}_{~~c}\left[\frac{C+\mu^{2}}{1-\mu^{2}}L^{c}_{x}(x;\lambda)-\frac{C+\lambda^{2}}{1-\lambda^{2}}L^{L,c}_{x}(x;\mu)\right]\delta(x-y) \nonumber \\
&-&\frac{\lambda+\mu}{(1-\lambda^{2})(1-\mu^{2})}\gamma^{ab}\partial_{x}\delta(x-y)\,. \label{3.12}
\end{eqnarray}
With the tensor product notation, it can be rewritten as 
\begin{eqnarray}
\left\{L^{L}_{x}(x;\lambda),\otimes L^{L}_{x}(y;\mu)\right\}_{\rm P}
&=& \left[r^{L}(\lambda,\mu),L^{L}_{x}(x;\mu)\otimes 1+1\otimes L^{L}_{x}(x;\mu)\right]\delta(x-y) \nonumber \\
&& -\left[s^{L}(\lambda,\mu),L^{L}_{x}(x;\mu)\otimes 1-1\otimes L^{L}_{x}(x;\mu)\right]\delta(x-y) \nonumber \\
&& -2s^{L}(\lambda,\mu)\partial_{x}\delta(x-y)\,. 
\end{eqnarray}
Here classical $r$-matrix $r^{L}(\lambda,\mu)$ and $s$-matrix $s^{L}(\lambda,\mu)$ 
are defined as\footnote{In \cite{KYhybrid}, the classical integrability is discussed by following \cite{Duncan}. 
On the other hand, we have followed the formalism in \cite{Maillet}. This is a new result.}
\begin{eqnarray}
&&r^{L}(\lambda,\mu) \equiv \frac{1}{2\left(\lambda-\mu\right)}\left(\frac{C+\mu^{2}}{1-\mu^{2}}+\frac{C+\lambda^{2}}{1-\lambda^{2}}\right)\left(T^{+}\otimes T^{-}+T^{-}\otimes T^{+}+T^{3}\otimes T^{3}\right)\,, \nonumber \\
&&s^{L}(\lambda,\mu) \equiv \frac{\lambda+\mu}{2\left(1-\lambda^{2}\right)\left(1-\mu^{2}\right)}\left(T^{+}\otimes T^{-}+T^{-}\otimes T^{+}+T^{3}\otimes T^{3}\right)\,.
\end{eqnarray}
The $r$/$s$-matrices are of rational type and satisfy the extended classical Yang-Baxter equation,
\begin{eqnarray}
&&\left[(r+s)^{L}_{13}(\lambda,\nu),(r-s)^{L}_{12}(\lambda,\mu)\right]
+\left[(r+s)^{L}_{23}(\mu,\nu),(r+s)^{L}_{12}(\lambda,\mu)\right] \nonumber \\
&& \qquad +\left[(r+s)^{L}_{23}(\mu,\nu),(r+s)^{L}_{13}(\lambda,\nu)\right]=0\,, 
\end{eqnarray}
where the subscripts denote the vector spaces on which the $r$/$s$-matrices act. 
Thus the classical integrability has been shown in the rational description. 

\section{Trigonometric description}

It is a turn to consider the classical integrability in the trigonometric description based on $U(1)_{\rm R}$\,.  
The $U(1)_{\rm R}$ symmetry is realized as the isometry of the target space and so 
the corresponding current is obtained by the Noether procedure as  
\begin{eqnarray}
j^{R,3}_{\mu}=2(1+C){\rm Tr}\left(T^{3}J_{\mu}\right)\,.
\label{norm}
\end{eqnarray}
The normalization is taken for later convenience. 

The $SU(2)_{R}$ symmetry of round $S^3$ is broken due to the squashing. However it is still possible to 
find out non-local conserved currents for the broken components, 
\begin{eqnarray}
j^{R,\pm}_{\mu} = 
2\,{\rm e}^{\gamma\chi}
\left(\eta_{\mu\nu} 
\pm i\sqrt{C}\epsilon_{\mu\nu}\right){\rm Tr}\left(T^{\pm}J^{\nu}\right)\,.
\label{non-local}
\end{eqnarray}
Here $\gamma$ is related to the squashing parameter $C$ like 
\begin{eqnarray}
\gamma \equiv \frac{\sqrt{C}}{1+C}\,. 
\end{eqnarray}
The scalar field $\chi(x)$ is defined as 
\begin{eqnarray}
\chi(x) \equiv -\frac{1}{2}\int\!\!dy\,\epsilon(x-y)\,j^{R,3}_{t}(y)\,,
\end{eqnarray}
where $\epsilon(x-y)$ is the sign function 
\begin{eqnarray}
\epsilon(x-y)\equiv
\left\{
\begin{array}{ccc}
+1 & \quad {\rm for} & x>y \\
-1 & \quad {\rm for} & x<y \\
\end{array}
\right.\,. 
\end{eqnarray}
Thus $\chi$ is non-local and so the currents in (\ref{non-local}) are also non-local. 
Note that the boundary condition (\ref{bc}) ensures the convergence of the above integral 
for an arbitrary value of $x$ and the following relation is useful,  
\begin{eqnarray}
\partial^{\mu}\chi=-\epsilon^{\mu\nu}j^{R,3}_{\nu}\,.
\label{derivative chi}
\end{eqnarray}
To show the conservation of (\ref{non-local}) directly,  
it is necessary to use the relations (\ref{eompm}) and (\ref{derivative chi}). 

As a matter of course, the Noether charge 
\begin{eqnarray}
Q^{R,3}=\int^{\infty}_{-\infty}\!\!\!dx~j^{R,3}_{t}(x)
\end{eqnarray}
generates the right multiplication of $T^{3}$\,, 
\begin{eqnarray}
\delta^{R,3}g=\left\{g,Q^{R,3}\right\}_{\rm P}=-gT^{3}\,.
\end{eqnarray}
Similarly, non-local charges 
\begin{eqnarray}
Q^{R,\pm}=\int^{\infty}_{-\infty}\!\!\!dx~j^{R,\pm}_{t}(x)
\end{eqnarray}
generate non-local transformations 
\begin{eqnarray}
\delta^{R,\pm}g 
= \{g,Q^{R,\pm}\}_{\rm P}
= -g\left[T^{\pm} {\rm e}^{\gamma\,\chi} - \gamma T^{3}\xi^{\pm}
\right]\,. \label{non-local trans}
\end{eqnarray}
Here new non-local fields have been introduced 
\begin{eqnarray}
\xi^{\pm}(x) \equiv -\frac{1}{2}\int\!\!dy\,\epsilon(x-y)\,j^{R,\pm}_{t}(y)\,. 
\end{eqnarray}
Note that $\xi^{\pm}$ are well defined under the boundary conditions (\ref{bc}) like $\chi$\,. 
For $C=0$, the non-local transformations (\ref{non-local trans}) reduce to the right multiplication of $T^{\pm}$ of $SU(2)_{\rm R}$\,. 

One can directly check that the action (\ref{sigma}) is invariant under the non-local transformations (\ref{non-local trans}).
It is necessary to use the equations of motion (\ref{eom}) to show the invariance and hence 
the non-local transformations generate ``on-shell'' symmetries. 

The Poisson brackets of $j^{R,\pm}_{t}$ and $j^{R,3}_{t}$ are 
\begin{eqnarray}
\left\{j^{R,\pm}_{t}(x),\,j^{R,\mp}_{t}(y)\right\}_{\rm P}&=& \mp i\,{\rm e}^{2\gamma\,\chi(x)}\,
j^{R,3}_{t}(x)\delta(x-y) \nonumber \\
&=&\pm \,\frac{i}{2\gamma}\partial_{x}
\left[{\rm e}^{2\gamma\,\chi(x)}\right]\delta(x-y)\,, \nonumber \\
\left\{j^{R,\pm}_{t}(x),\,j^{R,\pm}_{t}(y)\right\}_{\rm P} 
&=& \pm i\, \gamma\,\epsilon(x-y)\,j_{t}^{R,\pm}(x)j_{t}^{R,\pm}(y)\,, \nonumber  \\ 
\left\{j^{R,\pm}_{t}(x),\,j^{R,3}_{t}(y)\right\}_{\rm P} 
&=& \pm i\,j^{R,\pm}_{t}(x)\delta(x-y)\,. \nonumber 
\end{eqnarray}
By using 
\[
\chi(\pm\infty)=\mp Q^{R,3}/2
\] 
the classical Poisson brackets of the charges are evaluated as  
\begin{eqnarray}
\left\{Q^{R,+},Q^{R,-}\right\}_{\rm P} =-i\,\frac{q^{Q^{R,3}}-q^{-Q^{R,3}}}{q-q^{-1}}\,, \qquad 
\left\{Q^{R,\pm},Q^{R,3}\right\}_{\rm P} = \pm i\, Q^{R,\pm}\,. 
\label{q-deformed}
\end{eqnarray}
Here a constant parameter 
\begin{eqnarray}
q\equiv {\rm e}^{\gamma} =\exp\left(\frac{\sqrt{C}}{1+C}\right)
\end{eqnarray}
has been introduced 
and $Q^{R,\pm}$ are rescaled as 
\begin{eqnarray}
Q^{R,\pm} \rightarrow \left(\frac{\gamma}{\sinh\gamma} \right)^{1/2}
\, Q^{R,\pm}\,.
\end{eqnarray}
The algebra (\ref{q-deformed}) is nothing but a $q$-deformation of the $SU(2)_R$ Lie algebra \cite{Drinfeld,Jimbo}. 
The normalization of (\ref{norm}) is fixed so that the expression of the second commutator in (\ref{q-deformed}) is obtained. 

Then let us consider the following Lax pair, which is given in \cite{FR},  
\begin{eqnarray}
&&L^{R}_{t}(x;\lambda)=-\frac{1}{2}\sum_{a=1}^{3}\left[w_a(\lambda + \alpha)J^a_+(x) + w_a(\lambda - \alpha)J^a_-(x)\right]T^a\,, \nonumber \\
&&L^{R}_{x}(x;\lambda)=-\frac{1}{2}\sum_{a=1}^{3}\left[w_a(\lambda + \alpha)J^a_+(x) - w_a(\lambda - \alpha)J^a_-(x)\right]T^a\,, \nonumber \\
&&J_{\pm}=J_{t}\pm J_{x}\,,\quad C=-\tanh^{2}\alpha\,.
\label{right lax}
\end{eqnarray}
Here $\lambda$ is a spectral parameter and $w_a(\lambda)$ are defined as 
\begin{eqnarray}
w_1(\lambda) = w_2(\lambda) \equiv \frac{\sinh\alpha}{\sinh\lambda}\,, \quad w_3(\lambda) \equiv \frac{\tanh\alpha}{\tanh\lambda}\,. \nonumber
\end{eqnarray}
By definition, $\alpha$ can take a complex value, while $C$ must be real. 
Therefore $\alpha$ should be real or purely imaginary. 
For $\alpha = i\beta$~($\beta$:~real)\,, $C = \tan^2\,\beta$\,. 
Then the range of $C$ is naturally restricted to the physical region $C \geq -1$\,. 
By scaling $\lambda$ as $\lambda = \alpha \tilde{\lambda}$ 
and taking the $\alpha \to 0$ limit in (\ref{right lax}), 
the Lax pair of rational type for $SU(2)_{\rm R}$ is reproduced. 

The commutation relation
\begin{eqnarray}
\left[\partial_{t}-L^{R}_{t}(\lambda),\partial_{x}-L^{R}_{x}(\lambda)\right]=0 \nonumber
\end{eqnarray}
leads to the equations of motion (\ref{eom}) with the help of the flatness of $J=g^{-1}dg$\,.
Then the monodromy matrix is constructed with (\ref{right lax}) as 
\begin{eqnarray}
U^{R}(\lambda)\equiv{\rm P}\exp\left[\int^{\infty}_{-\infty}\!\!dx\,L^{R}_{x}(x;\lambda)\right]\,, \nonumber
\end{eqnarray}
and it is conserved,
\begin{eqnarray}
\frac{d}{dt}U^R(\lambda)=0\,. \nonumber 
\end{eqnarray}

The Poisson brackets of the $L^{R,a}_{x}(x;\lambda)$ are evaluated as 
\begin{eqnarray}
\left\{L^{R,\pm}_{x}(x;\lambda),L^{R,\mp}_{x}(y;\mu)\right\}_{\rm P}
&=&\mp if(\lambda,\mu)\left[h(\mu)L^{R,3}_{x}(x;\lambda)-h(\lambda)L^{R,3}_{x}(y;\mu)\right]\delta(x-y) \nonumber \\
&&-f(\lambda,\mu)\left[h(\mu)-h(\lambda)\right]\partial_{x}\delta(x-y)\,, \nonumber \\
\left\{L^{R,\pm}_{x}(x;\lambda),L^{R,3}_{x}(y;\mu)\right\}_{\rm P}
&=&\pm if(\lambda,\mu)\left[g(\lambda,\mu)h(\mu)L^{R,\pm}_{x}(x;\lambda)-h(\lambda)L^{R,\pm}_{x}(y;\mu)\right]\delta(x-y)\,, \nonumber \\
\left\{L^{R,3}_{x}(x;\lambda),L^{R,3}_{x}(y;\mu)\right\}_{\rm P}
&=&-f(\lambda,\mu)g(\lambda,\mu)\left[h(\mu)-h(\lambda)\right]\partial_{x}\delta(x-y)\,. \label{lax_poisson}
\end{eqnarray}
Here $f(\lambda-\mu)$, $g(\lambda-\mu)$, and $h(\lambda)$ are defined as:
\begin{eqnarray}
&& f(\lambda,\mu)\!\equiv\frac{1}{\sinh{(\lambda-\mu)}}\,,\qquad g(\lambda,\mu)\equiv\cosh(\lambda-\mu)\,, \nonumber \\
&& h(\lambda)\equiv\frac{\sinh{\alpha}\cosh{\alpha}\sinh^{2}{\lambda}}{\sinh{(\lambda-\alpha)}\sinh{(\lambda+\alpha)}}\,.
\end{eqnarray}
With the tensor product notation, it can be rewritten as follows:
\begin{eqnarray}
\left\{L^{R}_{x}(x;\lambda),\otimes L^{R}_{x}(y;\mu)\right\}_{\rm P}
&=& \left[r^{R}(\lambda,\mu),L^{R}_{x}(x;\mu)\otimes 1+1\otimes L^{R}_{x}(x;\mu)\right]\delta(x-y) \nonumber \\
&& -\left[s^{R}(\lambda,\mu),L^{R}_{x}(x;\mu)\otimes 1-1\otimes L^{R}_{x}(x;\mu)\right]\delta(x-y) \nonumber \\
&& -2s^{R}(\lambda,\mu)\partial_{x}\delta(x-y)\,. 
\end{eqnarray}
Here the classical $r$-matrix $r^{L}(\lambda,\mu)$ and $s$-matrix $s^{L}(\lambda,\mu)$ 
are given by 
\begin{eqnarray}
&&r^{R}(\lambda,\mu) \equiv \frac{h(\mu)+h(\lambda)}{2\sinh{\left(\lambda-\mu\right)}}\left(T^{+}\otimes T^{-}+T^{-}\otimes T^{+}+\cosh{\left(\lambda-\mu\right)}T^{3}\otimes T^{3}\right)\,, \nonumber \\
&&s^{R}(\lambda,\mu) \equiv \frac{h(\mu)-h(\lambda)}{2\sinh{\left(\lambda-\mu\right)}}\left(T^{+}\otimes T^{-}+T^{-}\otimes T^{+}+\cosh{\left(\lambda-\mu\right)}T^{3}\otimes T^{3}\right)\,.
\end{eqnarray}
The $r$/$s$-matrices are of trigonometric type. 
It is easy to show the extended classical Yang-Baxter equation is satisfied,
\begin{eqnarray}
&&\left[(r+s)^{R}_{13}(\lambda,\nu),(r-s)^{R}_{12}(\lambda,\mu)\right]
+\left[(r+s)^{R}_{23}(\mu,\nu),(r+s)^{R}_{12}(\lambda,\mu)\right] \nonumber \\
&& \qquad +\left[(r+s)^{R}_{23}(\mu,\nu),(r+s)^{R}_{13}(\lambda,\nu)\right]=0\,, 
\end{eqnarray}
where the subscripts denote the vector spaces on which the $r$/$s$-matrices act.

Finally let us comment on the equivalence between the two descriptions. 
The two Lax pairs in both descriptions lead to the identical equations of motion (\ref{eom}) 
and thus the two descriptions are equivalent as in principal chiral models. 
Reflecting this fact, there is a simple map between the flat $SU(2)_{\rm L}$ current $j^{L}_{\mu}$
and non-local, $q$-deformed $SU(2)_{\rm R}$ current $j^{R}_{\mu}$. 
\begin{eqnarray}
j_{\mu}^{R,\pm} = 2\, {\rm e}^{\gamma\chi}\,{\rm Tr}(g^{-1}j_{\mu}^L gT^{\pm})\,,  \qquad 
j_{\mu}^{R,3} = 2\, {\rm Tr}(g^{-1}j_{\mu}^L g T^3)\,. 
\end{eqnarray}
The current circumstance is quite similar to the Seiberg-Witten map \cite{SW}. 
The topological term in the left $SU(2)$ current $j^{L}_{\mu}$ may be regarded as a constant two-form flux 
and the existence of $q$-deformed $SU(2)_{\rm R}$ symmetry implies a ``quantum space'' such as a noncommutative space. 

\section{Conclusion and Discussion}

In this short summary we have considered the classical integrable structure of non-linear sigma models 
on three-dimensional squashed spheres. The classical integrability has been shown in the two descriptions, 
1) the rational description and 2) the trigonometric one. 

The rational description is related to the $SU(2)_{\rm L}$ symmetry. 
The improved  $SU(2)_{\rm L}$ conserved current contains a topological term so that it satisfies the flatness condition. 
Then the Lax pair is constructed with the improved current. 
It leads to the rational $r$/$s$-matrices satisfying the extended classical Yang-Baxter equation. 

On the other hand, the trigonometric description is related to the $U(1)_{\rm R}$ symmetry. 
It is enhanced to $q$-deformed $SU(2)_{\rm R}$ as a non-local symmetry. 
The Lax pair in this description leads to the trigonometric $r$/$s$-matrices. 
Thus the squashed sigma models can be described both in the rational description and trigonometric description. 
One may call this feature ``hybrid classical integrability''. 
The two descriptions are equivalent due to the presence of non-local map. 

As discussed in \cite{KY}, the $SU(2)_{\rm L}$ symmetry is enhanced to Yangian symmetry. 
The charges are obtained from the monodromy matrix in the rational description, $U^{L}(\lambda)$. 
Similarly, the $q$-deformed $SU(2)_{\rm R}$ symmetry is enhanced to a quantum affine algebra \cite{KMY}. 
The charges are also constructed by expanding $U^{R}(\lambda)$. It would be interesting to consider an affine extension 
of $q$-deformed Poincare symmetry in Schr\"odinger sigma models \cite{KYSch}. 

Our analysis here is applicable to squashed Wess-Zumino-Novikov-Witten models. In fact,  
it is shown that the $SU(2)_{\rm L}$ symmetry is enhanced to the Yangian symmetry \cite{KOY} 
even in the presence of Wess-Zumino term. Thus one may expect that the classical dynamics is described 
as a rational model in the description based on $SU(2)_{\rm L}$\,. 
It is a future problem to study the classical integrability in the description based on $U(1)_{\rm R}$\,. 

Another issue is to construct the Bethe ansatz based on the integrable structure discussed here. 
The resulting Bethe equations should be a composite of XXX and XXZ models for the left and right. 
In fact, quantum solutions are already known \cite{quantum1,quantum2,quantum3}, 
though the classical integrable structure we revealed here has not been discussed there. 
It would be interesting to consider them in the context of AdS/CFT.

\section*{Acknowledgment}
The work of IK was supported by the Japan Society for the Promotion of Science (JSPS). 
The work of KY was supported by the scientific grants from the Ministry of Education, Culture, Sports, Science 
and Technology (MEXT) of Japan (No.\,22740160). 
This work was also supported in part by the Grant-in-Aid 
for the Global COE Program ``The Next Generation of Physics, Spun 
from Universality and Emergence'' from 
MEXT, Japan.

\end{document}